\documentstyle[prc,preprint,tighten,aps]{revtex}
\begin {document}
\draft
\title{
Three-body resonances in \bbox{^6}He, \bbox{^6}Li, and \bbox{^6}Be,
and the soft dipole mode problem of neutron halo nuclei}
\author{Attila Cs\'ot\'o\cite{email}}
\address{W.~K. Kellogg Radiation Laboratory, 106-38, California
Institute of Technology,\\Pasadena, California 91125, USA}
\date{\today}

\maketitle

\begin{abstract}
\noindent
Using the complex scaling method, the low-lying three-body
resonances of $^6$He, $^6$Li, and $^6$Be are
investigated in a parameter-free microscopic three-cluster
model. In $^6$He a 2$^+$, in $^6$Li a 2$^+$ and a 1$^+$, and
in $^6$Be the 0$^+$ ground state and a 2$^+$ excited state
is found. The other experimentally known 2$^+$ state of
$^6$Li cannot be localized by our present method. We have
found no indication for the existence of the predicted 1$^-$
soft dipole state in $^6$He. We argue that the sequential
decay mode of $^6$He through the resonant states of its
two-body subsystem can lead to peaks in the excitation
function. This process can explain the experimental results
in the case of $^{11}$Li, too. We propose an experimental
analysis, which can decide between the soft dipole mode
and the sequential decay mode.
\end{abstract}

\narrowtext

\section{Introduction}

Recently, some experimental and theoretical analyses claimed the
discovery of a new type of collective excitations, the soft
dipole mode, in nuclei far from stability. It seems probable
now, that in certain nuclei, near the neutron drip line, a dilute
neutron halo with large spatial extension can exist around a
compact core \cite{Tanihata,Kobayashi1,Kobayashi2,Hansen,Niigata};
e.g.\ in $^{11}$Li and
$^6$He two neutrons form the neutron halo around the $^9$Li
and $^4$He cores, respectively. Shortly after their discovery
it was predicted that the oscillation of the halo neutrons against
the core can produce collective excited states in these nuclei
\cite{Ikeda}. These collective states are thought to be similar
to the giant dipole resonances, where the neutrons of a nucleus
oscillate against its protons. In the neutron halo nuclei the
restoration force is weak (soft), which leads to low-frequency,
viz.\ low-energy (1-5 MeV) states compared to the giant dipole
resonances ($\sim$ 20 MeV).

The overwhelming majority of the experimental and theoretical
studies confirm the existence of these low-energy soft dipole
modes. However, all these works without exception conclude for
the existence of these states from the behavior of certain
quantities (bound state approximated energies and electric dipole
strengths \cite{EDS}, excitation cross sections
\cite{Kobayashi1,Kobayaship,Sakuta,Brady}, dipole sum rules
\cite{Suzuki}, etc.) at real energies, i.e.\ they are all
indirect evidences of the soft
dipole mode, thus they are far from being unambigous. For
instance, recently we demonstrated \cite{3br} that the possibility of
resonance+scattering type asymptotic behaviors in a three-body
system can lead to an apparent, resonance-like structure in the
three-body continuum. To show up a direct proof for the existence
of the soft dipole mode, we must find three-body resonant poles
of the scattering matrix at compex energies with the predicted
properties (excitation energy, decay width, spin-parity).

In this paper we search for three-body resonances in $^6$He
using the complex scaling method. In order to test our method,
we study the $^6$Li and $^6$Be nuclei, as well.

\section{Model}

Our model is a microscopic three-cluster ($\alpha +N+N$) approach
to the six-nucleon systems. The trial function of the six-body
problem has the form
\widetext
\begin{equation}
\Psi =\sum_{(i,j,k)}\sum_{S,l_1,l_2,L}
{\cal A}\left \{\left [ \left [\Phi^i(\Phi ^j\Phi^k)
\right ]_S \chi ^{i(jk)}_{[l_1l_2]L}(
\mbox{\boldmath $\rho $}_1,
\mbox{\boldmath $\rho $}_2)\right ]_{JM}\right \},
\label{wfn}
\end{equation}
\narrowtext
\noindent
where the $i,j$, and $k$ indices denote any one of the labels $\alpha$,
$n$, and $p$, and the first sum runs over all possible different
arrangements of the clusters [$\alpha (nn)$ and $n(\alpha n)$ in the
case of $^6$He, $\alpha (pn)$, $n(\alpha p)$, and $p(\alpha n)$
in the case of $^6$Li, and $\alpha (pp)$ and $p(\alpha p)$ in the
case of $^6$Be].
In (\ref{wfn}) ${\cal A}$ is the intercluster antisymmetrizer,
the $\Phi$ cluster internal states are translation invariant
harmonic oscillator shell model states, the \mbox{\boldmath
$\rho $} vectors are the different intercluster Jacobi
coordinates, and [\ ] denotes angular momentum coupling.
In the sum over $S,l_1,l_2$, and $L$ we should include all
angular momentum configurations of any significance.

It is of prime importance to choose a nucleon-nucleon
interaction which is appropriate for all subsystems
appearing in the model space \cite{He6}. We choose the
Minnesota interaction \cite{Tang}, containing central,
Coulomb, and spin-orbit terms,with the parameters as in
\cite{He6}. It was pointed out \cite{He6}, that using the
size parameter $\beta =0.606 fm^{-2}$ for the $\alpha$
particle, which minimizes its internal energy, together
with the exchange mixture parameter $u=0.98$ of the
Minnesota force, the experimental $\alpha +N$ phase shifts
can be excellently reproduced (see Fig.\ 1 of \cite{He6}).
This force also gives very good phase shifts in the $^1$$S_0$,
and $^3$$P_J$ (J=0,1,2) states of the $n+n$ and $p+p$ systems.
Since our interaction cannot account for the charge independence
breaking of the $N-N$ force due to the different $\pi ^0$ and
$\pi^\pm$ pion masses, our $p+n$ phase shifts are slightly worse
than the $n+n$ and $p+p$ ones (see the discussion in \cite{He6}).
It was emphasized in \cite{He6,beta} that the Minnesota force is
inappropriate in the $^3$$S_1$ $N+N$ state, because it reproduces
the ground state of
the deuteron without the tensor coupling of the $^3$$D_1$ state.
This leads to a too strong interaction in the $^3$$S_1$ $N+N$
state. Fortunately, in $^6$He and $^6$Be a triplet-even state
cannot be present between the outer neutrons and protons,
respectively, thus this problem does not appear. In $^6$Li,
however, we have to pay definite attention to this problem.
We neglect the tensor force because there is no firm ground
to determine its parameters. For test purposes we shall use
the tensor force of Ref.\ \cite{Heiss,PRL}, which reproduces
the $J=1,2,0$ order of the low-energy $^3$$P_J$ phase shifts
\cite{He6}.

In a similar model \cite{He6} but using three size parameters
for the $\alpha$ (allowing two monopole excited states in addition
to the ground state) and also including rearrangement (3+3 nucleon)
channels, we got excellent results for the ground state of $^6$He
and its (bound state approximated) isobar analogues in $^6$Li
and $^6$Be. In that work we demonstrated that the high-lying
3+3 rearrangement channels are responsible for the 0.3-0.6 MeV
energy lack appearing in the three-body models of the $A=6$
nuclei (for a recent review of the three-body models see
\cite{Nico}, and the references
therein). The $^6$He wave function of \cite{He6} was used in the
study of the beta delayed deuteron emission from $^6$He, and
proved to give good agreement with the experiment \cite{beta}.
Unfortunately, the use of that huge model space would be hardly
managable computationally, and would cause numerical
instabilities in the present work. Using only one size parameter
for the $\alpha$ particle, and no rearrangements, we can keep
the excellent reproduction of the $\alpha +N$ phase shifts,
which means that the description of the dynamics of the two-body
subsystems is correct. The price, we have to pay for this model
space reduction, is that the two-neutron separation energy in the
ground state of $^6$He is not correct, being 0.670 MeV to be
compared to the experimental value 0.975 MeV. We could get closer
to the experiment including the $t+t$ rearrangement channel. However,
the inclusion of this channel into the $^6$He wave function require
the good reproduction of the $\alpha +n$ phase shifts in a
$\{ \alpha +n, t+d\}$ coupled channel model \cite{He6}. Although,
using three $\alpha$ size parameters we got excellent $\alpha +n$
phase shifts in such a model \cite{He6}, using only
one $\alpha$ size parameter it is no longer possible. The source
of the problem is probably the change of the threshold splitting
between the $\alpha +n$ and $d+t$ channels. We emphasize that
requiring the good reproduction of the $N+N$ and $\alpha +N$ data,
there is no remaining free parameter left in the model.

The next step is to find a method which can handle three-body
resonances. There are indirect and direct ways of doing that.
The indirect methods, e.g.\ \cite{Danilin}, study the
three-body problem at real energies, and extract resonance
parameters from the three-body phase shifts. In the direct methods
the aim is to find the complex energy poles of the three-body
scattering matrix. For example, in \cite{Matsui,Afnan,Emelyanov}
the authors determined the pole positions of
the three-body $S$-matrix by analytically continuing the homogenous
Faddeev-equation to complex energies. To get a decisive answer to
the question of the soft dipole mode of $^6$He, we must use a direct
method. Our choice is the complex scaling method (CSM) \cite{CSM}.
This method reduces the problem of resonant states to that of bound
states, and can handle the Coulomb interaction, which causes
enormous difficulties in the Faddeev framework, without any
problem. It has been demonstrated recently, that the CSM can be
used to find three-body resonances in such problems which are
very similar to our present situation \cite{3br}. Here we recall
only the main points of the CSM, the details can be found in
\cite{3br} and the references therein.

Suppose, we search for ($N$-body) resonant states of the Hamiltonian
$\widehat H$. We define a new Hamiltonian by
\begin{equation}
\widehat H_\theta=\widehat U(\theta)\widehat H\widehat U^{-1}
(\theta),
\end{equation}
where the $\widehat U(\theta)$ unbounded similarity transformation
acts, in the coordinate space, on a function $f({\bf r})$ as
\begin{equation}
\widehat U(\theta)f({\bf r})=e^{3i\theta/2}f({\bf r}e^{i\theta}).
\label{CS}
\end{equation}
(If $\theta$ is real, $\widehat U(\theta)$ means a rotation
into the complex coordinate plane, if it is complex, it means
a rotation and scaling.)
In the case of a many-body Hamiltonian, (\ref{CS}) means that the
transformation has to be performed in each Jacobi coordinate.
If the potential is dilation analytic \cite{Reed}, then there
is the following connection between the spectra of $\widehat H$
and $\widehat H_\theta$ \cite{ABC}:
(i) the bound eigenstates of $\widehat H$ are the
eigenstates of $\widehat H_\theta$, for any
value of $\theta$ within $0\leq\theta<\pi/2$;
(ii) the continuous spectrum of $\widehat H$ will be
rotated by an angle 2$\theta$;
(iii) the complex generalized eigenvalues of
$\widehat H_\theta$, $\varepsilon_{\rm res}=E
-i\Gamma /2$, E,$\Gamma >0$
(where $\Gamma$ is the full width at half maximum)
belong to its proper spectrum, with square-integrable
eigenfunctions, provided $2\theta >\vert \arg
\varepsilon_{\rm res}\vert$.
In nuclear physics the CSM has been successfully applied in the
RGM description of $^8$Be \cite{Be8}, in an OCM model of
$^{20}$Ne \cite{KruppaKato}, and in the OCM description of
the resonances of $^{10}$Li \cite{IkedaKato}.

Since the resonant wave functions become localized 
in the CSM, we can use any bound-state method to describe them.
Here we use an expansion of the wave function, in terms of
products of tempered Gaussian functions of the Jacobi
coordinates, in a variational method \cite{He6}. A
typical term of this expansion looks like $\rho_1^{l_1}
\exp [-(\rho_1/\gamma_i)^2]Y_{l_1m_1}(\widehat
\rho_1)\cdot \rho_2^{l_2}
\exp [-(\rho_2/\gamma_j)^2]Y_{l_2m_2}
(\widehat \rho_2)$,
where $l_1$ and $l_2$ are the angular momenta in the
two relative motions, respectively, and the widths
$\gamma$ of the Gaussians are the parameters of the expansion.
We choose ten $\gamma$ parameters in each relative motion,
and they follow a geometric progression \cite{Kamimura}
with $\gamma_1=1$ and $\gamma_{10}=15$.

\section{Results}

We have carried out calculations for the $0^+$, $1^+$, $2^+$,
$0^-$, $1^-$, and $2^-$ states in the $A=6$ nuclei. In order
to keep all important angular momentum channels in Eq.\
(\ref{wfn}), and at the same time keep the size of the problem
manageble, we should know the contribution of each channel
to the total wave functions. That is why we first made real
energy calculations (i.e., without complex scaling) for the
above-mentioned states, using square-integrable trial functions.
The resulting weights of the different orthogonal $(S,L)$
components are in Table \ref{tab1}. The number of different
$\{i(jk);S,(l_1l_2)L\}$ configurations in Eq.\ (\ref{wfn})
was typically 10-12, most of them are non-orthogonal to each other.
We calculated the amount of their clustering, which measure
the weights of the different non-orthogonal channels
\cite{Lovas,Li6,He6}, and kept the most important 5-6
channels in the complex calculations.

In $^6$Li we left out all angular momentum configuration which
contain triplet-even state between the outer proton and neutron.
The inclusion of the configurations which contain the $^3$$S_1$
$p+n$ state (e.g.\ $\{ \alpha (pn); S, (l_1l_2)L=1,(02)2\}$ in
the $2^+$ state of $^6$Li) would suppress all other channels
because, the Minnesota force is too strong in this $p+n$ state.
For example, this would lead to 99.8 \% weight of the $(S,L)=(1,2)$
component in the $2^+$ state of $^6$Li, while without this
$\{ \alpha (pn); 1,(02)2\}$ configuration the weight is only 0.5 \%,
as we can see in Table \ref{tab1}. With their omission we dropped only
a tiny part of the model space, because these components have
large overlaps with the non-orthogonal angular momentum
configurations in the $(\alpha p)n$ and $(\alpha n)p$ partitions
\cite{Li6}. Without a tensor force, the omission of the configurations
which contain the $^3$$D_1$ $p+n$ state, almost does not change
anything. If a tensor force and the $^3$$D_1$ $p+n$ components
(e.g.\ the $\{ \alpha (pn); 1,(20)2\}$ configuration in the $1^+$
state of $^6$Li) were present, the $^6$Li would become $\sim$10 MeV
overbound. This is because our force would lead to a $\sim$10 MeV
overbinding in the case of the $^3$$S_1-$$^3$$D_1$ coupled
deuteron, itself. Although, we have no explicit $^3$$S_1$ $p+n$
state in $^6$Li, the coupling of the $\{ \alpha (pn); 1,(20)2\}$
state to the $(S,L)=(1,0)$ channels, which are non-orthogonal to
the omitted $\{ \alpha (pn); 1,(00)0\}$ component, would result
in this $\sim$10 MeV overbinding of $^6$Li. With the omission of the
triplet-even $p+n$ components in $^6$Li, these problems do not appear,
but the $1^+$ ground state becomes $\sim$1 MeV underbind. So,
in conclusion, there is definitely room for improvements in $^6$Li.

The parameters of all three-body resonances
which we found are listed in Table \ref{tab2}, along with the
experimental energies and widths. We solved the complex scaled
Schr\"odinger equations with different $\theta$ rotation angles.
Without approximations, the resulting pole position would not
depend on $\theta$. Using an approximate method (in our case a
variational method), the pole position slightly depend on $\theta$,
and the stationary point of the $\theta$-trajectory should be
accepted as the position of the resonance \cite{Be8}. Our best
$\theta$ values were around 0.25-0.3 radian.

As a consequence of our not complete model space, which resulted
in the $\sim$1 MeV underbinding in the the ground state of $^6$Li,
the $1^+$ resonance is pushed to higher energy, considerably
increasing its width. The same underbinding effect which appeared
in the ground state of $^6$He pushes the slightly bound $0^+$
state of $^6$Li into the continuum, resulting a three-body
resonance. The experimentally known $2^+$ state of $^6$Li
at E=0.610 MeV with $\Gamma$=1.7 MeV cannot be resolved by
our present method because the pole at $(0.610-i0.85)$ MeV is
mixed up with the points of the rotated discretized continuum,
using any rotation parameter.

The agreement of the resonance parameters with the experimental
values is good for the $2^+$ states of $^6$He, $^6$Li, and $^6$Be,
and the $0^+$ state of $^6$Be. It is important to note that
while our interaction underbinds the $0^+$ state of $^6$He
by about 0.3 MeV, the $2^+$ state is ``overbound'' by about
0.1 MeV (and its width is smaller than the experimental value,
accordingly). It means that the sometimes usual way to refit the
interaction strength so as to get the correct binding energy for
the ground state is dangerous. In our present case it would result
in the parameters $E=0.46$ MeV and $\Gamma =0.008$ MeV for the $2^+$
state. The best thing we can do is to accept the results, coming
from an interaction which is good for the subsystems, as they are.
To test the effect of the tensor interaction, we performed a
calculation for the $2^+$ state of $^6$He with the tensor
force of \cite{Heiss,PRL} included. It resulted in $E=0.76$ MeV
and $\Gamma= 0.07$ MeV parameters, thus it has small effect.

To illustrate the working mechanism of our method, we show in
Fig.\ \ref{fig1} the results of the calculations for the $2^+$
and $1^+$ state in $^6$Li. As we can see, the rotated discretized
continuum is a mixture of the three-body scattering continuum
points, starting at the origin, and the resonance+scattering
continua, starting at the $3/2^-$ and $1/2^-$ resonance positions
of the $^5$He and $^5$Li subsystems, respectively \cite{3br}.
The probably not accurate enough double precision numerics and the
interaction between the continua, which are close to each other,
makes the picture a bit fuzzy, but the identification of the
resonant states is unambiguous.

In Fig.\ \ref{fig2} we show our
result for the $1^-$ state of $^6$He, where the soft dipole mode
was predicted at $(6-i2.5)$ MeV ($E=6$ MeV, $\Gamma =5$
MeV, \cite{Sakuta}). As we can see, our model does not confirm
the existence of such a state. No resonant state occured while
increasing the $\theta$ value to as large as 0.7, where the CSM
becomes unstable. This result is in agreement with \cite{strength}
and \cite{Ferreira}, where the authors did not find any indication
for the existence of a $1^-$ state, analyzing the $^6$He strength
functions and break-up cross sections at real energies.

In the literature there are works for the $A=6$ nuclei in which
the three-body resonance parameters are extracted from real
energy phase shifts \cite{Danilin,Hofmann}, and Faddeev
calculations \cite{Matsui,Afnan,Emelyanov}, respectively.
The results of these works are consistent with each other and
with ours, except that in \cite{Hofmann}, in addition to the
well-known $0^+$ ground state and $2^+$ excited state, the
authors found low-lying narrow $0^+$ and $1^+$ excited states
in $^6$Be, a fact which we try to explain later.

In \cite{3br} we argued that we can identify three-body
resonances using the fact that they have to appear in each
possible Jacobi coordinate system. Here, using the $2^+$ state
of $^6$He, we check this claim. Switching off all $\alpha (nn)$
and $n(\alpha n)$ channels, respectively, the resonance occurs
at $1.01-i0.081$ MeV and $1.89-i0.094$ MeV, respectively.
If we keep only the $(S,L)=(0,2)$ channels, we get the resonance
at $2.00-i0.446$ MeV, while keeping only the $(S,L)=(1,1)$
configurations the resonance appears at $2.04-i0.760$ MeV. Thus,
our present model shows that the three-body resonances appear not
just in every different Jacobi coordinate systems, but in every
different $(S,L)$ configuration. It means that a $1^-$ three-body
resonance, if exists, must show up even if our model space
were very restrictive.

To sum up our results, we have found the experimentally known
resonances, but not found any indication that the $1^-$ soft
dipole mode exists in $^6$He. In the next section we give a
possible explanation of the experimental finding of soft
dipole resonances in neutron halo nuclei.

\section{Discussion}

We have shown in \cite{3br} that if there are resonances in a
two-body subsystem of a three-body system, then the
sequentially decay of the system through these resonant
states leads to structures in the three-body continuum, in addition
to the structures coming from the three-body resonances.
This is because around the two-body resonance energy, in addition
to the three-body phase space, a substantial two-body phase
space is available for the system to decay into. We emphasize,
that these structures have kinematical, rather than dynamical origin.
The $0^+$ and $1^+$ excited states of $^6$Be, found in
\cite{Hofmann}, are nice examples to demonstrate the operation
of the sequential decay mode. The model wave function of
\cite{Hofmann} consists of $^3$He+$^3$He and $^5$Li+$p$
channels with, bound state approximated, $3/2^-$, $1/2^-$, and
$3/2^+$ $^5$Li states. The $0^+$ and $1^+$ resonances of
\cite{Hofmann} in the $^6$Be phase shifts are apparent. They
are the consequence of the fact that the $^5$Li+$p$ asymptotic decay
modes are built into the wave function. These states are sequential
decay modes, having two-body dynamical structures. Thus, these
are not three-body resonances in $^6$Be. In a similar work
\cite{Danilin} the method of the hyperspherical harmonics
does not allow sequential decay modes, that is why the authors
did not find excited $0^+$ and $1^+$ in the $^6$Be phase shifts.

We can conclude from our results that it is possible that the
soft dipole mode does not exist in $^6$He, and the experiments
see only the $^5$He+$n$ sequential decay mode of this nucleus.
We show that there is no experimental finding which contradicts
our assumption.

In $^6$He the sequential decay mode leads to a superposition
of the $3/2^-$ and $1/2^-$ $^5$He states and the neutron in the
final state. The angular momentum of the relative motion is
most probably $l=0$ to avoid the centrifugal barrier. As
a consequence, $0^-$, $1^-$, and $2^-$ structures can appear
in the excitation function of $^6$He. From the $0^+$ ground state,
only the $1^-$ state can be excited conventionally, the other
two states can only be excited by parity transfer e.g.\ by using
pions. Due to the splitting of the $3/2^-$ and $1/2^-$
states of $^5$He (these states are at 0.89 MeV and 4 MeV
\cite{Ajzenberg}, respectively), two $1^-$ bumps can appear
in the excitation function of $^6$He at 0.975+0.89 $\approx$ 1.9
MeV and 0.975+4 $\approx$ 5 MeV excitation energies, respectively
(0.975 MeV is the
two-neutron separation energy of $^6$He). In recent experimets
\cite{Sakuta,Brady} a $1^-$ structure appeared around 6 MeV, and
the well-known $2^+$ state at 1.8 MeV. The position of the $1^-$
bump is in good agreement with our prediction, while our other
structure at 1.9 MeV coincides with the position of the $2^+$
state. We can see in Fig.\ 2 of \cite{Sakuta} that the assumption
of a mixed $2^+$ and $1^-$ nature of the bump at 1.8 MeV is not
in contradiction with the measured angular distribution, on the
contrary, it would improve the agreement between theory
and experiment.

Let us point out that a similar situation may occur in $^{11}$Li.
Here the sequential decay mode leads to $^{10}$Li+$p$
final states. There are several measurements for the low-lying
resonances of $^{10}$Li (see e.g.\ \cite{Li10old}), the most recent
Ref.\ \cite{Li10}, which seems to be the most reliable, gives
a $1^+$ state at 0.42 MeV, and a $2^+$ state at 0.8 MeV
excitation energies. The most recent value of the two-neutron
separation energy of $^{11}$Li is 0.295$\pm$0.035 MeV
\cite{Li11}. The sequential decay mode can lead to structures
around 0.42+0.29 $\approx$ 0.7 and 0.8+0.29 $\approx$ 1.1 MeV
$^{11}$Li excitation energies. Assuming again that the $l=0$
angular momentum between $^{10}$Li and $n$ is preferred,
the possible $J^\pi$ values are $1/2^+$ and $3/2^+$ at 0.7 MeV,
and $1/2^+$, $3/2^+$, and $5/2^+$ at 1.1 MeV. All of these
$J^\pi$ states can be excited from the $3/2^-$ ground state of
$^{11}$Li without parity transfer, which means that the
experimental structure is a coherent superposition of them.
In \cite{Kobayashi1,Kobayaship} a bump occured in the
$^{11}$Li excitation spectrum around 1.2 MeV excitation energy.
The position and the experimentally observed $1/2^+$, $3/2^+$,
or $5/2^+$ spin-parity character of this state are again in good
agreement with our result. In \cite{Li10} a state in $^{10}$Li
is also found around 4 MeV, and in \cite{Kobayashi1,Kobayaship}
we can see some structures in this energy region, too.

In Ref.\ \cite{MSU} the authors studied the Coulomb break-up
of $^{11}$Li on Pb target. They found that the average velocity
of the outcoming $^9$Li nuclei is larger than the average
velocity of the outcoming neutrons. They explained this by the
post acceleration of the $^9$Li nuclei in the field of the lead
target nucleus. From the large velocity shift they concluded that
the breakup process should be direct, not resonant through the
presumed soft dipole mode. If the sequential decay mode causes
the apparent resonance-like structure in $^{11}$Li, as we assume,
then the post acceleration is felt by the $^{10}$Li particle (and
not by the other neutron), thus
the average neutron velocity can be smaller than that of the
$^9$Li velocity. In Fig.\ 14 of \cite{MSU} one can see two
peaks of the velocity shift curve. The larger shift is probably
due to the direct breakup (the three-body scattering final state),
while the other peak at much smaller shift can account for
the sequential decay.

The complete kinematical measurement of \cite{MSU} would allow
a stringent test of our assumption that there is no soft
dipole mode in $^{11}$Li, only a sequential decay mode. In the
center of mass frame one should see a considerable increase of
the 180 degree angular correlation between $^9$Li and
neutrons around the 1.2 MeV excitation of $^{11}$Li.

\section{Conclusions}

In summary, we have searched for low-energy three-body
resonances in $^6$He, $^6$Li, and $^6$Be, using the
complex scaling method in a microscopic three-cluster model.
Our model can account both for the correct nuclear physics and
the proper three-body dynamics. We have found the experimentally
known resonances, except a $2^+$ state of $^6$Li, which
cannot be localized by our present method. However, we
have not found the predicted $1^-$ soft dipole mode in $^6$He.
We argued that this state, if exists, should appear even in
a model which is much simpler than ours.

We concluded from our result, that it is possible that the soft
dipole modes of the neutron halo nuclei do not exist, and the
experiments see just the sequential decay modes of these nuclei.
We have shown, through the examples of $^6$He and $^{11}$Li that
there is no experimental fact which contradicts this assumption.
A test of our assumption could come from the analysis of the
experimental core--neutron angular correlation in break-up reactions.
If the sequential decay mode has a significant effect, then a
strong increase of the 180 degree correlation in the center of
mass frame should be found around the ``resonance'' energies.

Finally, we mention that the importance of the sequential
decay modes was pointed out in \cite{Tanihatap}, claiming
that the broad part of the  momentum distribution of the
projectile fragments in the neutron halo nuclei probably
comes from these modes.

\acknowledgments

This work was supported by a Fulbright Fellowship and NSF Grants
Nos. PHY90-13248 and PHY91-15574 (USA), and by OTKA Grants
Nos.\ 3010 and F4348 (Hungary).
I wish to thank S.~E. Koonin and K. Langanke for useful comments
on the manuscript.

\begin{figure}
\caption{Energy eigenvalues of the complex scaled Hamiltonian
of the (a) $2^+$, and (b) $1^+$ states of $^6$Li. The dots
are the points of the rotated discretized continua, while the
circles are three-body resonances. The rotation angle is 0.3 rad.}
\label{fig1}
\end{figure}

\begin{figure}
\caption{Energy eigenvalues of the complex scaled Hamiltonian
of the $1^-$ states of $^6$He. The rotation angle is 0.3 rad.}
\label{fig2}
\end{figure}

\mediumtext
\begin{table}
\caption{The weights (in percents) of the orthogonal $(S,L)$
components in the various $J^\pi$ states of $^6$He, $^6$Li,
and $^6$Be.}
\begin{tabular}{cc@{}r@{}lc@{}r@{}lc@{}r@{}lc@{}r@{}lc@{}r@{}lc@{}r@{}l}
 & \multicolumn{3}{c}{$0^+$}  & \multicolumn{3}{c}{$1^+$}  &
 \multicolumn{3}{c}{$2^+$}  & \multicolumn{3}{c}{$0^-$}  &
 \multicolumn{3}{c}{$1^-$}  & \multicolumn{3}{c}{$2^-$}  \\
\tableline
$^6$He & (0,0)&\  86.&5 & (0,1)&\    0.&5& (0,2)&\  52.&8&
         (1,1)&\ 100.&0 & (0,1)&\   86.&0& (0,2)&\ $<$0.&1 \\

       & (1,1)&\  13.&5 & (1,0)&\    2.&0& (1,1)&\  46.&5&
              &      &  & (1,1)&\   14.&0& (1,1)&\ 100.&0 \\

       &      &\     &  & (1,1)&\   97.&1& (1,2)&\   0.&7&
              &      &  & (1,2)&\ $<$0.&1& (1,2)&\ $<$0.&1 \\

       &      &\     &  & (1,2)&\    0.&4&      &\     & &
              &      &  &      &\      & &      &\     &  \\
\hline
$^6$Li & (0,0)&\  87.&3 & (0,1)&\    3.&3& (0,2)&\  50.&0&
         (1,1)&\ 100.&0 & (0,1)&\   81.&4& (0,2)&\ $<$0.&1 \\

       & (1,1)&\  12.&7 & (1,0)&\   96.&0& (1,1)&\  49.&5&
              &      &  & (1,1)&\   18.&6& (1,1)&\ 100.&0 \\

       &      &\     &  & (1,1)&\    0.&7& (1,2)&\   0.&5&
              &      &  & (1,2)&\ $<$0.&1& (1,2)&\ $<$0.&1 \\

       &      &\     &  & (1,2)&\ \tablenote{Not included.}
                                       & &      &\     & &
              &      &  &      &\      & &      &\     &  \\
\hline
$^6$Be & (0,0)&\  87.&7 & (0,1)&\    0.&2& (0,2)&\  59.&9&
         (1,1)&\ 100.&0 & (0,1)&\   84.&0& (0,2)&\ $<$0.&1 \\

       & (1,1)&\  12.&3 & (1,0)&\    0.&6& (1,1)&\  39.&6&
              &      &  & (1,1)&\   16.&0& (1,1)&\ 100.&0 \\

       &      &\     &  & (1,1)&\   99.&1& (1,2)&\   0.&5&
              &      &  & (1,2)&\ $<$0.&1& (1,2)&\ $<$0.&1 \\

       &      &\     &  & (1,2)&\    0.&1&      &      & &
              &      &  &      &       & &      &\     &  \\
\end{tabular}
\label{tab1}
\end{table}

\narrowtext
\begin{table}
\caption{Energies (relative to the $\alpha$ energy) and
full widths of three-body resonances in $^6$He, $^6$Li,
and $^6$Be. The experimental values are taken from
\protect\cite{Ajzenberg}.}
\begin{tabular}{ccr@{}lr@{}lr@{}lr@{}l}
 & & \multicolumn{4}{c}{Theory}  & \multicolumn{4}{c}{Experiment} \\
\cline{3-6} \cline{7-10}
 & & \multicolumn{2}{c}{E (MeV)}&\multicolumn{2}{c}{$\Gamma$ (MeV)}&
     \multicolumn{2}{c}{E (MeV)}&\multicolumn{2}{c}{$\Gamma$ (MeV)} \\
\tableline
$^6$He & $2^+$ & 0.&74 & 0.&06  & 0.822&$\pm$0.025 & 0.113&$\pm$0.020 \\
$^6$Li & $0^+$ & 0.&22 & 0.&001 & \multicolumn{2}{c}{$-$0.137
 \tablenote{Bound state.}}                         &      &           \\
       & $2^+$ &\multicolumn{4}{c}{\tablenote{Cannot be resolved by
the present method (see the text).}}
                                & 0.610&$\pm$0.022 & 1.7  &$\pm$0.2   \\
       & $2^+$ & 1.&59 & 0.&28  & 1.696&$\pm$0.015 & 0.54 &$\pm$0.020 \\
       & $1^+$ & 5.&71 & 3.&89  & 1.95 &$\pm$0.05  & 1.5  &$\pm$0.2   \\
$^6$Be & $0^+$ & 1.&52 & 0.&16  & 1.371&$\pm$\tablenote{No errors
                                 are given in \protect\cite{Ajzenberg}.}
                                                   & 0.092&$\pm$0.006 \\
       & $2^+$ & 2.&81 & 0.&87  & 3.04 &$\pm$0.05  & 1.16 &$\pm$0.06   \\
\end{tabular}
\label{tab2}
\end{table}


\begin{references}
\bibitem[*]{email} E-mail address: H988CSO@HUELLA.BITNET \\
On leave from: Institute of Nuclear Research of the Hungarian
Academy of Sciences, P.O.Box 51, Debrecen, H--4001, Hungary
\bibitem{Tanihata} I. Tanihata, Nucl. Phys. {\bf A553},
361c (1993).
\bibitem{Kobayashi1} T. Kobayashi, Nucl. Phys. {\bf A538},
343c (1992).
\bibitem{Kobayashi2} T. Kobayashi, Nucl. Phys. {\bf A553},
465c (1993).
\bibitem{Hansen} P.~G. Hansen, Nucl. Phys. {\bf A553}, 89c (1993).
\bibitem{Niigata} {\em Proceedings of the International Symposium
on Structure and Reactions of Unstable Nuclei} (Niigata, Japan,
1991), edited by K. Ikeda and Y. Suzuki (World Scientific,
Singapore, 1991).
\bibitem{Ikeda} P.~G. Hansen and B. Jonson, Europhys. Lett.
{\bf 4}, 409 (1987); K. Ikeda, in Ref.\ \protect\cite{Niigata},
p.\ 3; Y. Suzuki and K. Ikeda, Phys. Rev. C {\bf 38}, 410 (1988).
\bibitem{EDS} N. Poppelier, L. Wood, and P. Glaudemans, Phys. Lett.
{\bf B157}, 120 (1985); Y. Suzuki and Y. Toshaka, Nucl. Phys.
{\bf A517}, 599 (1990); A.~C. Hayes and D. Strottman, Phys. Rev.
C {\bf 42}, 2248 (1990).
\bibitem{Kobayaship} T. Kobayashi, in Ref.\ \protect\cite{Niigata},
p.\ 187.
\bibitem{Sakuta} S.~B. Sakuta, A.~A. Ogloblin, O.~Ya. Osadchy,
Yu.~A. Glukhov, S.~N. Ershov, F.~A. Gareev, and J.~S. Vaagen,
Europhys. Lett. {\bf 22}, 511 (1993).
\bibitem{Brady} F. Brady, G.~A. Needham, J.~L. Romero,
C.~M. Castaneda, T.~D. Ford, J.~L. Ullmann, and M.~L. Webb,
Phys. Rev. Lett. {\bf 51}, 1320 (1983).
\bibitem{Suzuki} Y. Suzuki, Nucl. Phys. {\bf A528}, 395 (1991).
\bibitem{3br} A. Cs\'ot\'o, CALTECH preprint MAP-165 (1993),
submitted to Phys. Rev. C.
\bibitem{He6} A. Cs\'ot\'o, Phys. Rev. C {\bf 48}, 165 (1993).
\bibitem{Tang} D.~R. Thompson, M. LeMere and Y.~C. Tang, Nucl.
Phys. {\bf A268}, 53 (1977); I. Reichstein and Y. C. Tang,
Nucl. Phys. {\bf A158}, 529 (1970).
\bibitem{beta} A. Cs\'ot\'o and D. Baye, Phys. Rev. C, January
(1994).
\bibitem{Heiss} P. Heiss and H.~H. Hackenbroich, Phys. Lett.
{\bf 30B}, 373 (1969).
\bibitem{PRL} A. Cs\'ot\'o, R.~G. Lovas, and A.~T. Kruppa,
Phys. Rev. Lett. {\bf 70}, 1389 (1993).
\bibitem{Nico} N.~W. Schellingerhout, L.~P. Kok, S.~A. Coon,
and R.~M. Adam, Phys. Rev. C {\bf 48}, 2714 (1993).
\bibitem{Danilin} B.~V. Danilin and M.~V. Zhukov,
Yad. Fiz. {\bf 56}, 67 (1993) [Phys. At. Nucl. {\bf 56},
460 (1993)].
\bibitem{Matsui} Y. Matsui Phys. Rev C {\bf 22}, 2591 (1980).
\bibitem{Afnan} A. Eskandarian and I.~R. Afnan, Phys. Rev. C
{\bf 46}, 2344 (1992).
\bibitem{Emelyanov} V.~G. Emelyanov, V.~I. Klimov, and
V.~N. Pomerantsev, Phys. Lett. {\bf 157B}, 105 (1985).
\bibitem{CSM} Y.~K. Ho, Phys. Rep. {\bf 99}, 1 (1983);
N. Moiseyev, P.~R. Certain, and F. Weinhold,
Mol. Phys. {\bf 36}, 1613 (1978); Proceedings of the Sanibel
Workshop Complex Scaling, 1978 [Int. J. Quantum Chem. {\bf 14},
343 (1978)]; B.~R. Junker, Adv. At. Mol. Phys. {\bf 18}, 207
(1982); W.~P. Reinhardt, Annu. Rev. Phys. Chem. {\bf
33}, 223 (1982); {\em Resonances--The Unifying Route Towards
the Formulation of Dynamical Processes, Foundations and
Applications in Nuclear, Atomic and Molecular Physics},
edited by E. Br\"andas and N. Elander, Lecture Notes in Physics
Vol. 325 (Springer-Verlag, Berlin, 1989).
\bibitem{Reed} M. Reed and B. Simon, {\em Methods of Modern
Mathematical Physics} (Academic Press, New York, 1978).
\bibitem{ABC} J. Aguilar and J.~M. Combes, Commun. Math. Phys.
{\bf 22}, 269 (1971); E. Balslev and J.~M. Combes, {\em ibid.}
{\bf 22}, 280 (1971); B. Simon, {\em ibid.} {\bf 27}, 1 (1972).
\bibitem{Be8} A.~T. Kruppa, R.~G. Lovas, and B. Gyarmati,
Phys. Rev. C {\bf 37}, 383 (1988).
\bibitem{KruppaKato} A.~T. Kruppa and K. Kat\= o, Prog. Theor.
Phys. {\bf 84}, 1145 (1990).
\bibitem{IkedaKato} K. Kat\=o and K. Ikeda, Prog. Theor. Phys.
{\bf 89}, 623 (1993).
\bibitem{Kamimura} H. Kameyama, M. Kamimura and M. Kawai,
in Ref.\ \protect\cite{Niigata}, p.\ 203.
\bibitem{Lovas} R. Beck, F. Dickmann and R. G. Lovas, Ann. Phys.
(N.~Y.) 173 (1987) 1.
\bibitem{Li6} A. Cs\'ot\'o and R.~G. Lovas, Phys. Rev. C {\bf
46}, 576 (1992).
\bibitem{Ajzenberg} F. Ajzenberg-Selove, Nucl. Phys. {\bf A490},
1 (1988).
\bibitem{strength} B.~V. Danilin, M.~V. Zhukov, J.~S. Vaagen,
and J.~M. Bang, Phys. Lett. B {\bf 302}, 129 (1993).
\bibitem{Ferreira} L.~S. Ferreira, E. Maglione, J.~M. Bang,
I.~J. Thompson, B.~V. Danilin, M.~V. Zhukov, and J.~S. Vaagen,
Phys. Lett. B {\bf 316}, 23 (1993).
\bibitem{Hofmann} H.~M. Hofmann and W. Zahn, Nucl. Phys.
{\bf A368}, 29 (1981).
\bibitem{Li10old} K.~H. Wilcox, R.~B. Weisenmiller, G.~J.
Wozniak, N.~A. Jelley, D. Ashery, and J. Cerny, Phys. Lett.
{\bf 59B}, 142 (1975); A.~I. Amelin, M.~G. Gornov, Yu.~B.
Gurov, A.~L. Ilin, P.~V. Morokhov, V.~A. Pechkurov, V.~I.
Savelev, F.~M. Sergeev, S.~A. Smirnov, B.~A. Chernyshev,
R.~R. Shafigullin, and A.~V. Shishkov, Yad. Fiz. {\bf 52},
1231 (1990) [Sov. J. Nucl. Phys. {\bf 52}, 782 (1990)].
\bibitem{Li10} H.~G. Bohlen, B. Gebauer, M. von Lucke-Petsch,
W. von Oertzen, A.~N. Ostrowski, M. Wilpert, Th. Wilpert,
H. Lenske, D.~V. Alexandrov, A.~S. Demyanova, E. Nikolskii,
A.~A. Korsheninnikov, A.~A. Ogloblin, R. Kalpakchieva,
Y.~E. Penionzhkevich, and S. Piskor, Z. Phys. A {\bf 344},
381 (1993).
\bibitem{Li11} B.~M. Young, W. Benenson, M. Fauerbach, J.~H.
Kelley, R. Pfaff, B.~M. Sherrill, M. Steiner, J.~S. Winfield,
T. Kubo, M. Hellstr\"om, N.~A. Orr, J. Stetson, J.~A. Winger,
and S.~J. Yennello, Phys. Rev. Lett. {\bf 71}, 4124 (1993).
\bibitem{MSU} D. Sackett, K. Ieki, A. Galonsky, C.~A. Bertulani,
H. Esbensen, J.~J. Kruse, W.~G. Lynch, D.~J. Morrissey,
N.~A. Orr, B.~M. Sherrill, H. Schulz, A. Sustich, J.~A. Winger,
F. De\'ak, \'A. Horv\'ath, \'A Kiss, Z. Seres, J.~J. Kolata,
R.~E. Warner, and D.~L. Humphrey, Phys. Rev. C {\bf 48}, 118 (1993).
\bibitem{Tanihatap} I. Tanihata, in Ref.\ \protect\cite{Niigata},
p.\ 233.
\end{references}
\end{document}